\documentstyle[psfig,11pt]{article}
\def\A{{\cal A}}

\def\Et{E^{\rm tot}}
\renewcommand{\Re}{\mbox{Re}}
\def\veps{\epsilon}

\def\r{\vec{r}}
\def\R{\vec{R}}
\def\G{{\vec{G}}}
\def\F{\vec{F}}

\def\k{\vec{k}}
\def\K{{\vec{K}}}

\def\S{\vec{S}}

\def\ba{\begin{array}}
 \def\ea{\end{array}}
\def\bc{\begin{center}}
 \def\ec{\end{center}}
\def\be{\begin{equation}}
 \def\ee{\end{equation}}
\def\bea{\begin{eqnarray}}
 \def\eea{\end{eqnarray}}
\def\bi{\begin{itemize}}
 \def\ei{\end{itemize}}
\def\bn{\begin{enumerate}}
 \def\en{\end{enumerate}}
\def\bF{\begin{figure}}
 \def\eF{\end{figure}}
\def\bt{\begin{tabular}}
 \def\et{\end{tabular}}
\def\bT{\begin{table}}
 \def\eT{\end{table}}

\def\>{$\rightarrow$}
\def\=>{$\Rightarrow$}
\def\nn\\{\nonumber \\}

\def\AQC#1{Advances in Quantum Chemistry {\bf #1}}

\def\CP#1{Commun.\,Phys.\,{\bf #1}}
\def\CPC#1{Comp.\,Phys.\,Commun.\,{\bf #1}}
\def\CPL#1{Chem.\,Phys.\,Lett.\,{\bf #1}}
\def\IJQCS#1{Int.\,J. Quantum.\,Chem. Suppl.\,{\bf #1}}

\def\JIMA#1{J.\,Inst.\,Maths.\,Appl.\,{\bf #1}}

\def\JMP#1{J.\,Math.\,Phys.\,{\bf #1}}

\def\JPCM#1{J.\,Phys.\,Condens.\,Matter\,{\bf #1}}
\def\JPCS#1{J.\,Phys.\,Chem.\,Solids\,{\bf #1}}
\def\JPF#1{J.\,Phys.\,F\,{\bf #1}}

\def\JVSTA#1{J.\,Vac.\,Sci.\,Technol.\,A\,{\bf #1}}
\def\MCP#1{Meth.\,Comp.\,Phys\,{\bf #1}}
\def\MP#1{Mol.\,Phys.\,{\bf #1}}
\def\PKM#1{Phys.\,Kondens.\,Mater.\,{\bf #1}}

\def\PR#1{Phys.\,Rev.\,{\bf #1}}

\def\PRA#1{Phys.\,Rev.\,A~{\bf #1}}
\def\PRB#1{Phys.\,Rev.\,B~{\bf #1}}
\def\PRL#1{Phys.\,Rev.\,Lett.\,{\bf #1}}

\def\SS#1{Surf.\,Sci.\,{\bf #1}}
\def\SSC#1{Solid State Commun.\,{\bf #1}}
\def\SSP#1{Solid State Phys.\,{\bf #1}}
\def\ZNF#1{Z.\,Naturforsch.\,{\bf #1}}
\def\ZPB#1{Z.\,Phys.\,B~{\bf #1}}

\begin{document}

\title{
{\vspace*{-1cm}
\flushleft
\ {\normalsize\tt To be published in Comp.\,Phys.\,Commun.\,}
       \vspace{1cm}\\}
Force calculation and atomic-structure optimization for
the full-potential linearized augmented plane-wave code
{\tt WIEN\rm}}
\author{\large Bernd Kohler,
Steffen Wilke, and Matthias Scheffler\\
\small\em Fritz-Haber-Institut der Max-Planck-Gesellschaft,\\
\small\em Faradayweg 4-6, D-14\,195 Berlin (Dahlem), Germany\\
\large\rm
Robert Kouba and Claudia Ambrosch-Draxl\\
\small\em Institut f\"ur theoretische Physik, Universit\"at Graz,\\
\small\em Universit\"atsplatz 5, A-8010 Graz, Austria
}
\date{submitted June 8, 1995}
\maketitle
\begin{abstract}{\small\noindent
Following the approach of Yu, Singh, and Krakauer
[Phys.\,Rev.\,B\,{\bf43} (1991) 6411]
we extended the linearized augmented plane wave code
\verb|WIEN| of Blaha, Schwarz, and coworkers by the
evaluation of forces.
In this paper we describe the approach, demonstrate
the high accuracy of the force calculation,  and use them for
an efficient geometry optimization of poly-atomic systems.
}\end{abstract}

\twocolumn
\newpage
\section*{PROGRAM SUMMARY}
{\footnotesize\sloppy\baselineskip 10pt
\ \\
{\em Title of program extension:} \verb|fhi95force|\\[1ex]
{\em catalogue number:} ... \\[1ex]
{\em Program obtainable from:}
 CPC Program Library, Queen's University of Belfast, N. Ireland
(see application form in this issue)\\[1ex]
{\em CPC Program Library programs used:}
{\em cat. no.:} ABRE;
{\em title:}    \verb|WIEN|;
{\em ref. in CPC:} 59 (1990) 399\\[1ex]
{\em Licensing provisions:} none \\[1ex]
{\em Computer, operating system, and installation:}
\bi
\item IBM RS/6000; AIX; Fritz-Haber-Institut
 der Max-Planck-Gesellschaft; Berlin.
\item CRAY Y-MP; UNICOS; IPP der Max-Planck-Gesellschaft; Garching.
\ei
{\em Operating system:} UNIX\\[1ex]
{\em Programming language:} FORTRAN77\\
(non-standard feature is the use of \verb|ENDDO|)\\[1ex]
{\em floating point arithmetic:} 64\,bits\\[1ex]
{\em Memory required to execute with typical data:}\\
64\,Mbyte (depends on case)\\[1ex]
{\em No.~of bits in a word:} 64\\[1ex]
{\em No.~of processors used:} one\\[1ex]
{\em Has the code been vectorized?} no\\[1ex]
{\em Memory required for test run:} ~64\,MByte\\[1ex]
{\em Keywords} \\
density functional theory, linearized augmented plane wave
method, LAPW, supercell, total energy, forces, structure
optimization, molecular dynamics, crystals, surfaces,
molecules\\[1ex]
{\em Nature of the physical problem}\\
For {\it ab-initio} studies of the electronic and
magnetic properties of poly-atomic systems, such
as molecules, crystals, and surfaces, it is
of paramount importance to determine stable and metastable
atomic geometries.
This task of structure optimization is greatly accelerated
and, in fact, often only feasible if the forces
acting on the atoms are known. \\
The computer-code described in this article enables
such calculations.
\\[1ex]
{\em Method of solution}\\
The full-potential linearized augmented plane wave (FP-LAPW)
method is well known to enable accurate calculations of
the electronic structure and magnetic properties of crystals
\cite{SkoelXX,SandeXX,Swimm81,Sjans84,Smatt86,Sblah90,Sblah93a,Ssing94}.
Within the supercell approach it has also been used for
studies of defects in the bulk and for crystal surfaces.
For the evaluation of the atomic forces within this method
we follow the approach outlined by Yu and coworkers
\cite{Syu91}.
In order to minimize the total energy as a function of
atomic positions we employ a damped Newton dynamics
scheme \cite{Sstum94} or alternatively the
variable metric algorithm of Broyden {\it et al.}
\cite{Sbroy73,Sbrod77,Sdenn83}.
Several applications of this approach to chemisorption at
surfaces have already been published \cite{Skohl95,Swilk95}.
\\[1ex]
{\em Restrictions on the complexity of the problem}\\
Inversion and orthorombic symmetry of the elementary cell is required.
\\[1ex]
{\em Typical running time}\\
The additional force calculation increases the
running time of a typical self-consistent
total energy calculation by 5-10\%.\\[1ex]
{\renewcommand{\,}{$\!$\ }
\renewcommand{\refname}{\ }
{\em References}\\
}}

\newpage
\onecolumn
\section*{LONG WRITE-UP}

\section{Introduction}
The augmented plane wave (APW) methods
\cite{slat37,slat64,louc67,matt68,dimm71}
and in particular its linearized form, the LAPW
\cite{brosXX,marc67,koelXX,andeXX,wimm81,jans84,matt86,blah90,sing94},
enable accurate calculations of the electronic and magnetic
properties of poly-atomic systems from first principles.
One successful implementation is the program package \verb|WIEN|.
This full-potential LAPW (FP-LAPW) code developed by Blaha, Schwarz
and coworkers \cite{blah90} has been successfully applied to
a wide range of problems \cite{blah85,kohl94a}
and systems such as complex crystals \cite{blah93b},
transition metal surfaces \cite{kohl95}, and molecules
(see the H$_2$ test case in this paper).

The main output of the \verb|WIEN| code is the total energy for
a given atomic arrangement.
Using only this quantity the minimization of the total energy of a
poly-atomic system is a costly and often impractible task.
However, the situation is changed if the forces which act on the
different atoms are available.
Only recently, force formulations within the LAPW method have
been introduced and tested by several authors
\cite{soleXX,yu91,yu92,goed92,krim94,faeh95}.
We followed the approach of Yu, Singh, and Krakauer (YSK)
\cite{yu91} and implemented the direct calculation of
atomic forces into the original \cite{blah90} and the
\verb|WIEN93| version \cite{blah93a} of the \verb|WIEN| code.
The obtained forces are highly accurate and can be used in an
efficient minimization scheme to optimize the geometry of
poly-atomic systems.

The remainder of the paper is organized as follows.
In Sec.\,II we summarize the most important features of
the YSK force formalism. Section\,III describes the
energy minimization procedure.
In Sec.\,IV results of test calculations are presented, and,
finally, Sec.\,V and VI describe the structure and the installation
of our program package \verb|fhi95force|.

\section{Evaluation of Forces}
\subsection{Forces within Density-Functional Theory}
Within density-functional theory the ground state total
energy is given by the minimum of a total energy
functional with respect to the electron density $n(\r)$
\be\label{efun}
\Et[n] = T[n] + U[n] + E^{\rm xc}[n]
\ee
where $T[n]$, $U[n]$, and $E^{\rm xc}[n]$
represent the functionals of the non-interacting many electron kinetic,
the electrostatic, and the exchange-correlation (xc) energy, respectively.
The electron density $n^{\rm sc}(\r)$ which  minimizes $\Et[n]$ is found
by solving self-consistently the Kohn-Sham (KS)
equations \cite{hohe64,kohn65}
\be\label{KS}
H \psi_i(\r) =
[\hat{T}+V^{\rm eff}(\r)] \psi_i(\r) =
\veps_i \psi_i (\r)
\ee
where $\hat{T}$ is the single-particle kinetic energy operator.
Throughout the paper we use Rydberg atomic units.
The effective potential $V^{\rm eff}(\r)$ is given by
\be
\label{effr}
V^{\rm eff} (\r)
=  V^{\rm es}(\r) + V^{\rm xc}(\r)
\ee
where
\be
V^{\rm es}(\r)
=\int d^3\r\,'\, \frac{n(\r\,')}{|\r-\r\,'|}
-\sum_I\frac{Z_I}{| \r-\R_I |}
\ee
denotes the total electrostatic potential
created by the electron density
\be\label{rho}
n(\r)=\sum_{i=1}^{\infty} f_i\psi_i^*(\r)\psi_i(\r)
\ee
and the nuclear charges.
The quantities $f_i$ are the occupation numbers of the eigenstates
$\psi_i(\r)$.
The $I$-th nucleus is positioned at $\R_I$ and carries the charge $Z_I$.
The xc potential $V^{\rm xc}(\r)$ is
\be
\label{potxcr}
V^{\rm xc}(\r)=
\frac{\delta E^{\rm xc}[n]}{\delta n}\quad.
\ee
The KS total energy $\Et[n]$ is then calculated
by using the expressions
\bea
\label{ekin} T[n]  \nonumber
&=& \sum_{i=1}^\infty f_i \veps_i - \int d^3\r\,n(\r) V^{\rm eff}(\r) \\
\label{ees} U[n]   \nonumber
&=& \frac{1}{2} \int d^3\r\,d^3\r\,'\,\frac{n(\r)n(\r\,')}{|\r-\r\,'|}
- \int d^3\r\,n(\r)\sum_I\frac{Z_I}{|\r-\R_I|}
+ \frac{1}{2}\sum_{I\neq J} \frac{Z_I Z_J}{|\R_I-\R_J|}\\
\label{exc} E^{\rm xc}[n]
&=& \int d^3\r\, n(\r) \veps^{\rm xc}\left[n\right](\r)
\quad,
\eea
where $\veps^{\rm xc}[n]$ is the xc energy per particle.
For finite temperatures, or in order to stabilize the convergence of
the self-consistent calculation the electronic states may be occupied
according to a Fermi distribution at a non-zero electron
temperature $T^{\rm el}$
\be
f_i=\left[1+\exp\left(
\frac{\veps_i-\veps_{\rm F}}{k_{\rm B}T^{\rm el}}
\right)\right]^{-1}
\ee
where $\veps_{\rm F}$ and $k_{\rm B}$ are the Fermi energy and the
Boltzmann constant. In this case, one has to minimize the free energy
\cite{gill89,neug92}
\be
{\cal F} [n]=\Et[n]-T^{\rm el}S^{\rm el}
\ee
with the entropy $S^{\rm el}$ given by
\be
S^{\rm el} =
-2 k_{\rm B} \sum_i
\left[
f_i \log f_i + (1-f_i) \log(1-f_i)\right] -S_0
\quad.
\ee
Here, $S_0$ is chosen such that the entropy $S^{\rm el}$ vanishes
for $T^{\rm el}=0$\,K.
The force on the $I$-th nucleus is defined as the negative
derivative of the free energy with respect to the nuclear
coordinate $\R_I$:
\be\label{Ffree}
\F_I
=-\left.\frac{d\Et - T^{\rm el}dS^{\rm el}}{d\R_I}
\right|_{n(\r)=n^{\rm sc}(\r)}
\quad.
\ee
It is evaluated by displacing the respective nucleus
by a small amount $\Delta \R_I$ and calculating the resulting
first-order change of the free energy $\Delta {\cal F}$.
For the different energy terms in eq.\,(\ref{exc}) we obtain
the following first order variations
\bea\nonumber
\Delta T[n]
&=& \sum_{i=1}^\infty \Delta f_i \veps_i
+ \sum_{i=1}^\infty f_i \Delta \veps_i
- \int d^3\r\,\Delta n(\r) V^{\rm eff}(\r)
- \int d^3\r\, n(\r) \Delta V^{\rm eff}(\r)\\
\label {eesdr}\Delta U[n] \nonumber
&=& \int d^3\r\, \Delta n(\r) V^{\rm es}(\r)
- \F_I^{\rm HF}\Delta \R_I \nonumber \\
\label{dxcr} \Delta E^{\rm xc}[n]
&=& \int d^3\r\, \Delta n(\r)  V^{\rm xc}(\r)
\quad.
\eea
The first term $\sum_i \Delta f_i \veps_i$  in
$\Delta T[n]$ is canceled with the contribution from the variation
of the entropy $S^{\rm el}$ in eq.\,(\ref{Ffree})
\cite{wein92}.
The Hellmann-Feynman (HF) force $\F_I^{\rm HF}$ describes
the classical electrostatic force exerted on the $I$-th nucleus by
all the other charges of the system (electrons and nuclei)
\cite{hell37,feyn39} and is obtained from
\be\label{FHF}
{\F}_I^{\rm HF}=Z_I \nabla_{\r}V_{I}^{\rm es}(\r)|_{\r=\R_I}
\ee
where $V_{I}^{\rm es}(\r)$ is the electrostatic potential
\be
V_{I}^{\rm es}(\r)=V^{\rm es}(\r)+\frac{Z_I}{|\r-\R_I|}
\ee
felt by the $I$-th nucleus.
Taking the definition of the effective potential  in eq.\,(\ref{effr})
into account the force on the $I$-th nucleus is then given by
\be\label{dEtot}
\F_I =\F_I^{\rm HF}
- \frac{1}{\Delta\R_I}\left\{
\sum_{i=1}^\infty f_i \Delta \veps_i
- \int d^3\r\, n^{\rm sc}(\r)\Delta V^{\rm eff}(\r)\right\}\quad.
\ee
It should be mentioned that the free energy variation
$\Delta {\cal F}$ and thus the forces $\F_I$ are invariant
to any first order deviation $\Delta n(\r)$ from the
self-consistent density $n^{\rm sc}(\r)$.
Thus, force expressions different from eq.\,(\ref{dEtot}) may be
derived if this variational freedom is used, e.g. if the
electron density is shifted rigidly with the nuclei
\cite{pett76,ande85,meth93}.

\subsection{Basis Set Corrections to the Hellmann-Feynman Force}
Usually, the HF force $\F_I^{\rm HF}$ can be evaluated quite easily
using eq.\,(\ref{FHF}).
The second term in eq.\,(\ref{dEtot}) describes the so-called
Pulay forces \cite{pula69}.
The explicit expression of this correction to the HF force
$\F_I^{\rm HF}$ depends on how the KS equation is solved.
One usually expands a KS wavefunction $\psi\equiv\psi_i(\r)$ at the
eigenvalue $\veps\equiv\veps_i$ linearly using a set of basis
functions $\phi_\nu$:
\be
\psi = \sum_\nu C_\nu \phi_\nu
\quad.
\ee
With this variational basis functions, the KS equation
becomes the following secular equation
\be\label{secular}
\sum_\mu
\left(H_{\mu\nu}-\veps O_{\mu\nu}\right)C_\nu = 0\quad,
\ee
or equivalently
\be\label{secularB}
\sum_{\mu\nu}
C^*_\mu\left(H_{\mu\nu}-\veps O_{\mu\nu}\right)C_\nu = 0\quad.
\ee
The Hamilton and overlap matrix elements are defined as
\bea
H_{\mu\nu}&=&\langle \phi_\mu|H|\phi_\nu\rangle\\
O_{\mu\nu}&=&\langle \phi_\mu|\phi_\nu\rangle\quad.
\eea
Both eqs.\,(\ref{secular}) and (\ref{secularB}) also hold
for a shifted atomic configuration ${\R_I+\Delta \R_I}$.
Thus, we obtain for the latter
\be\label{secular_dR}
\sum_{\mu\nu}
\left(C^*_\mu+\Delta C^*_\mu\right)
\left[
H_{\mu\nu}+\Delta H_{\mu\nu}
- (\veps+\Delta\veps)(O_{\mu\nu}+\Delta O_{\mu\nu})
\right]
\left(C_\nu+\Delta C_\nu\right) = 0
\quad.
\ee
The variations of the matrix elements are as follows:
\bea
\Delta O_{\mu\nu}
&=& \langle\Delta\phi_\mu|\phi_\nu\rangle
+ \langle\phi_\mu|\Delta \phi_\nu\rangle
= 2 \Re \langle\Delta\phi_\mu|\phi_\nu\rangle\\
\Delta H_{\mu\nu}
&=& 2 \Re \langle\Delta\phi_\mu|H|\phi_\nu\rangle
+ \langle\phi_\mu|\Delta \hat{T}+\Delta V^{\rm eff}|\phi_\nu\rangle
\quad.
\eea
If we allow only first order changes in $\Delta\R_I$ and
take into account eq.\,(\ref{secular}) we arrive at the following
simplified form of eq.\,(\ref{secular_dR})
\be\label{secular_final}
\sum_{\mu\nu}C^*_\mu C_\nu
\left(\Delta H_{\mu\nu}
-\veps\Delta O_{\mu\nu}-\Delta\veps\,O_{\mu\nu} \right)=0
\quad.
\ee
Employing the  normalization conditions
\be
\sum_{\mu\nu} C^*_\mu C_\nu O_{\mu\nu} = 1
\ee
eq.\,({\ref{secular_final}) may be transformed to an expression
for the linear change $\Delta \veps$ of the KS eigenvalue $\veps$.
We obtain
\bea \nonumber
\Delta \veps_i
&=& \sum_{\mu\nu} C^*_\mu C_\nu
\left[\Delta H_{\mu\nu} -\veps_i\Delta O_{\mu\nu}\right] \\
&=& \label{deps}
\sum_{\mu\nu} C^*_\mu C_\nu \left[
2 \Re\langle\Delta\phi_\mu|H-\veps_i|\phi_\nu\rangle
+ \langle\phi_\mu|\Delta \hat{T} + \Delta V^{\rm eff}|\phi_\nu\rangle
\right]
\eea
and finally rewrite the second part of eq.\,(\ref{dEtot}):
\be\label{Ffinal}
\F_I=\F_I^{\rm HF}
- \sum_{i\mu\nu} f_i C^*_{i\mu} C_{i\nu}
\left[ 2 \Re
\langle\frac{d\phi_\mu}{d\R_I}|H-\veps_i|\phi_\nu\rangle
+ \langle\phi_\mu|\frac{d \hat{T}}{d \R_I}|\phi_\nu\rangle
 \right]
\ee
\quad.
Note that we use the relation
\be
\sum_{i\mu\nu} f_i C^*_{i\mu} C_{i\nu}
\langle\phi_\mu| \Delta V^{\rm eff}|\phi_\nu\rangle
= \int d^3\r\, n(\r) \Delta V^{\rm eff} (\r)\quad.
\ee
In eq.\,(\ref{Ffinal}) the first term within the square
brackets is called incomplete basis
set correction \cite{sataXX,bend83,sche85}.
Its existence was first noted by Hurley \cite{hurl54}.
The respective sum vanishes if the basis functions are
independent on the atomic positions or if their first order
changes $\Delta\phi_\mu$
lie completely within the subspace described by the
original basis set $\{\phi_\eta\}$, i.e.,
\be
| \Delta\phi_\mu \rangle
=\sum_\eta |\phi_\eta\rangle\langle\phi_\eta| \Delta\phi_\mu \rangle
\quad.
\ee
Then, eq.\,(\ref{secularB}) can be applied to eq.\,(\ref{Ffinal}) and
the incomplete basis set correction vanishes explicitly.
This is for example the case if the basis set $\{\phi_\eta\}$
is complete.
The second term of the HF force correction
$\langle\phi_\mu|d \hat{T}/d\R_I|\phi_\nu\rangle$
may be non-zero if the kinetic energy is position dependent,
e.g. if due to the use of a mixed basis set the calculated kinetic
energy is discontinuous.

Up to now the formulation for the total force
$\F_I$ has remained completely general. In the following we
will focus on the application of the outlined formalism within
the FP-LAPW method.

\subsection{LAPW Method}
In the augmented plane-wave (APW) methods space is divided
into the interstitial region (IR) and non-overlapping muffin-tin
(MT) spheres centered at the atomic sites \cite{slat37}.
By this the atomic-like character of the wavefunctions, potential,
and electron density close to the nuclei can be described
accurately as can be the smoother
behavior of these quantities in between the atoms.
In the IR the basis set consists of plane waves $\exp(i\K\r)$.
The choice of a computationally efficient and accurate representation
of the wavefunctions within the MT spheres has been discussed
by several authors, e.g. \cite{louc67,brosXX,marc67,andeXX}.
In the original APW formulation introduced by Slater
\cite{slat37,slat64} the plane-waves are augmented to the exact
solutions of the Schr\"odinger equation within the MT at the
calculated eigenvalues.
This approach is exact but computationally very expensive
because it
leads to an explicit energy dependence of the Hamilton
and overlap matrices.
Instead of performing a single diagonalization to solve the KS equation
one repeatedly needs to evaluate the determinant of the secular
equation (\ref{secular}) in order to find its zeros and thus the
single particle eigenvalues $\veps_i$.

In the linearized APW the difficulty is removed by using a fixed
set of suitable MT radial functions \cite{marc67,andeXX,brosXX}.
Within Andersen's approach, used also in the \verb|WIEN| code,
radial solutions $u_l^I(\veps_l^I,r_I)$ of the KS equation
at fixed energies $\veps_l^I$ and their energy derivatives
$\dot{u}_l^I(\veps_l^I,r_I)$ are employed.
Basically, this choice corresponds to a linearization of
$u_l^I(\veps,\r)$ around $\veps_l^I$ \cite{andeXX}.
The concept implies that the radial functions
$u_l^I(\veps_l)$ and $\dot{u}_l^I(\veps_l)$ and
the respective overlap and Hamilton matrix elements need
to be calculated only for a few energies $\veps_l^I$.
Moreover, all KS energies $\veps_i$ for one $\k$-point are found by
a single diagonalization (for a detailed discussion see \cite{sing94}).

The LAPW basis functions $\phi_{\K}(\r)$ which are used for the
expansion of the KS wavefunctions
\be
\psi_{\k,i}(\r) = \sum_{|\K|\leq K^{\rm wf}} C_{i}(\K) \phi_\K(\r)
\ee
are defined as
\be\label{wfbas}
\phi_{\K}(\r) =
\left\{ \ba{ll}
\Omega^{-1/2} \exp(i\K\r),& \r \in {\rm IR} \\
{\displaystyle \sum_{lm}[
a_{lm}^I(\K)u_l^I(\veps_l^I,r_I) +
b_{lm}^I(\K)\dot{u}_l^I(\veps_l^I,r_I)
]Y_{lm}(\hat{r}_I)},& r_I \le s_I \quad.
\ea\right.
\ee
Here, $\K=\k+\G$ denotes the sum of a reciprocal lattice vector
$\G$ and a vector $\k$ within the first Brillouin zone.
The wave function cutoff $K^{\rm wf}$ limits the number of
these $\K$ vectors and thus the size of the basis set.
The symbols in eq.\,(\ref{wfbas}) have the following meaning:
$\Omega$ is the unit cell volume, $s_I$ is the MT radius, and
$\r_I = \r-\R_I$ is a vector within the MT sphere of the $I$-th atom.
Note that $Y_{lm}(\hat{r})$ represents a complex spherical
harmonic with $Y_{l-m}(\hat{r})=(-1)^mY^*_{lm}(\hat{r})$.
The radial functions $u_l(\veps_l,r)$ and
$\dot{u}_l(\veps_l,r)$ are solutions of the equations
\bea\label{urad}
H^{\rm sph}\,u_l(\veps_l,r)Y_{lm}(\hat{r}) &=&
\veps_l\, u_l(\veps_l,r)   Y_{lm}(\hat{r})\\
\label{udot}
H^{\rm sph}\,\dot{u}_l(\veps_l,r) Y_{lm}(\hat{r}) &=&
 [\veps_l \dot{u}_l(\veps_l,r) +
u_l(\veps_l,r)] Y_{lm}(\hat{r})
\eea
which are regular at the origin. The operator
$H^{\rm sph}$ contains only the spherical average, i.e., the
$l=0$ component, of the effective potential within the MT.
The $\veps_l$ should be chosen somewhere within that energy band with
$l$-character.
By requiring that value and slope of the basis functions are
continuous at the surface of the MT sphere the coefficients
$a_{lm}(\K)$ and $b_{lm}(\K)$ are determined.

The representation of the potentials and densities resembles
the one employed for the wave functions, i.e.,
\be\label{vbas}
V^{\rm eff} (\r)=
\left\{\ba{ll}
\displaystyle
\sum_{|\G|\leq G^{\rm pot}} V_\G^{\rm eff} \exp(i\G\r),
& \r \in {\rm IR} \\
\displaystyle\hspace*{3.5mm}
\sum_{lm} V^{\rm eff}_{lm,I}(r_I) Y_{lm}(\hat{r}_I),&
r_I \le s_I\quad.
\ea\right.
\ee
Thus, no shape approximation is introduced. The quality of this
full-potential description is controlled by the cutoff parameter
$G^{\rm pot}$ for the lattice vectors $\G$ and the size of the
$(l,m)$-representation inside MTs.

\subsection{LAPW Forces}
The basis functions $\phi_\K(\r)$ defined in eq.\,(\ref{wfbas})
are centered at the nuclei positions $\R_I$ and thus move with the
atoms.
Furthermore, the single-particle kinetic energy is not continuous
at the MT sphere boundaries where both types of basis functions are
matched.
Thus, an accurate force formalism has to deal with
both matrix elements
\bea
\langle \frac{d \phi_{\K}}{d \R_I}|
H-\veps_i| \phi_{\K'}\rangle
&\mbox{and}&
\langle \phi_{\K}|\frac{d \hat{T}}{d \R_I}|\phi_{\K'}\rangle
\eea
in the correction to the HF force in eq.\,(\ref{dEtot}).
YSK derived expressions for both terms \cite{yu91,sing94}.
Independently, a successful formulation of the forces
was found by Soler and Williams \cite{soleXX,sole93}
who started from the kinetic energy functional
$T[n]=-\sum_i f_i \int d^3\r\,\psi^*_i \nabla^2 \psi_i$
and employed a formulation of the potential energy $U[n]$
introduced by Weinert \cite{wein81}.

In the following we briefly summarize the YSK method.
The force on the $I$-th atom can be written as
\be
\F_I\label{fa}
= \F_I^{\rm HF} + \F_I^{\rm core} + \F_I^{\rm semi} + \F_I^{\rm val}
\ee
where $\F_I^{\rm val}$ ($\F_I^{\rm semi}$) combines the
Pulay corrections due to valence (semicore) electrons
while $\F_I^{\rm core}$ denotes the respective core term.
The different contributions are
\bea
\F_I^{\rm HF} &=&
Z_I\sum_{m=-1}^1 \lim_{r_I\rightarrow 0}\,
\frac{V^{\rm es}_{1m,I} (r_I)}{r_I}\,
\nabla_{\r_I} [r_I Y_{1m} (\hat{r}_I)]\\
\F_I^{\rm core} &=& \label{fcor}
-\int d^3\r\, n^{\rm core}(\r) \nabla_{\r} V^{\rm eff}(\r) \\
\F_I^{\rm val} &=&  \label{fval}
 \int_I  d^3\r\, V^{\rm eff}(\r) \nabla_{\r} n^{\rm val}(\r)
 + \sum_{\k,i} f_{\k,i} \sum_{\K,\K'} C_i^*(\K')C_i(\K)\times\\
&& \nonumber
\mbox{ }\times \left[
(\K^2-\veps_i)\oint d\S_I\,\phi_{\K'}^*(\r)\phi_\K^{}(\r)
- i(\K-\K')\langle \phi_{\K'}|H-\veps_i|\phi_\K \rangle_I
\right]\quad.
\eea
The semicore correction $\F_I^{\rm semi}$ is equivalent to
eq.~\ref{fval}.
The evaluation of the HF force $\F_I^{\rm HF}$ is straight
forward in a FP-LAPW calculation because the electrostatic potential
$V_I^{\rm es}(\r)$ is needed already for the evaluation of
the KS effective potential $V^{\rm eff}(\r)$.
Hence, we obtain
\be \label{fhf}
\F_I^{\rm HF} =
Z_I \sqrt{\frac{3}{8\pi}} \lim_{r_I \rightarrow 0}
\frac{1}{r_I} \left(\ba{c}\displaystyle
-V^{\rm es}_{11,I} (r_{I})
+ V^{\rm es}_{1-1,I} (r_{I}) \\ \displaystyle
-iV^{\rm es}_{11,I} (r_{I})
- iV^{\rm es}_{1-1,I} (r_{I}) \\ \displaystyle
\sqrt{2}\,V^{\rm es}_{10,I} (r_{I})
\ea\right)\quad.
\ee
Alternatively to the approach of YSK, the core correction
$\F_I^{\rm core}$ in eq.\,(\ref{fcor}) can also be deduced
via eq.\,(\ref{Ffinal}).
Within the \verb|WIEN| code the core electron density
$n^{\rm core} (\r)$ is calculated using only the spherical part
of the Hamiltonian.
Hence, the core wavefunctions of the KS equation can be viewed
as an (incomplete) set of spherical basisfunctions
$\phi_\nu^{\rm core}(\r)$.
The derivative of these functions with respect to the atomic
position $\R_I$ is given by
\be
\nabla_{\R_I} \phi_\nu^{\rm core}(\r)
= -  \nabla_{\r} \phi_\nu^{\rm core}(\r)\quad.
\ee
Thus, the relevant matrix elements in the incomplete basis set
correction in eq.\,(\ref{Ffinal}) can be written as
\bea
\Re \langle -\nabla_{\r}
\phi_\nu^{\rm core}|H-\veps_i|\phi_\nu^{\rm core}\rangle
&=&
\int d^3\r\,\phi_\nu^{\rm core\ *} (\r)
\nabla_{\r} \left(H-\veps_i\right) \phi_\nu^{\rm core} (\r)\\
&& \label{Icor}
\mbox{ }-\int d^3\r\,\nabla_{\r} \left\{ \phi_\nu^{\rm core\ *} (\r)
\left[H-\veps_i\right] \phi_\nu^{\rm core} (\r) \right\}
\eea
which leads to eq.\,(\ref{fcor}) if we take into account that
$\nabla_{\r} (H-\veps_i) = \nabla_{\r} V^{\rm eff}(\r)$ and choose
the integration boundaries for the integral in eq.\,(\ref{Icor})
at the MT sphere boundaries where the functions $\phi_\nu^{\rm core}$
vanish.

The terms
$\langle \phi_{\K'}|H-\veps_i|\phi_\K \rangle_I$ in eq.\,(\ref{fval})
are given by the overlap and Hamilton matrix elements.
Naturally, the sum
$\sum_{\k,i}$ has to be executed after the determination of
the KS eigenvalues and the occupation numbers $f_{\k,i}$.
We are left with the integrals in eqs.\,(\ref{fcor}) and (\ref{fval}).
They can be derived from the general case [see eq.\,(A5) in YSK]
\bea \label{int}\nonumber
\lefteqn{\int d^3\r\, \nabla_{\r}
[f(r) Y_{lm}^*(\hat{r})]
 g(r) Y_{l'm'}(\hat{r}) =} \\
 & & \label{int1}  \int d^3\r\,
\frac{f(r)g(r)}{r} r\nabla_{\r} Y_{lm}^*(\hat{r}) Y_{l'm'}(\hat{r})\\
 & & \nonumber   \mbox{}+
\frac{d f(r)}{dr}g(r) \hat{r}\,Y_{lm}^*(\hat{r}) Y_{l'm'}(\hat{r})
\eea
which is calculated using for the first term on the right-hand side
of eq.\,(\ref{int1}):
\bea
r\left(\frac{d}{dx}+i\frac{d}{dy}\right) \,Y_{lm}(\hat{r}) &=&
+l    \,\sqrt{\frac{(l+m+1)(l+m+2)}{(2l+1)(2l+3)}} \,Y_{l+1,m+1}(\hat{r})
\nn\\
&&\mbox{}
+(l+1)\,\sqrt{\frac{(l-m)  (l-m-1)}{(2l+1)(2l-1)}} \,Y_{l-1,m+1}(\hat{r})
\nn\\
r\left(\frac{d}{dx}-i\frac{d}{dy}\right) \,Y_{lm}(\hat{r}) &=&
-l    \,\sqrt{\frac{(l-m+1)(l-m+2)}{(2l+1)(2l+3)}} \,Y_{l+1,m-1}(\hat{r})
\nn\\
&&\mbox{}
-(l+1)\,\sqrt{\frac{(l+m)  (l+m-1)}{(2l+1)(2l-1)}} \,Y_{l-1,m-1}(\hat{r})
\nn\\
r\frac{d}{dz} \,Y_{lm}(\hat{r}) &=&
- l     \,\sqrt{\frac{(l+m+1)(l-m+1)}{(2l+1)(2l+3)}} \,Y_{l+1,m}(\hat{r})
\nn\\
&&\mbox{}
+ (l+1) \,\sqrt{\frac{(l+m)  (l-m)}  {(2l+1)(2l-1)}} \,Y_{l-1,m}(\hat{r})
\quad.
\eea
The spherical integrals in the second part of eq.\,(\ref{int1}) and
in the surface integral in the second line of eq.\,(\ref{fval})
can be evaluated by transforming them into Gaunt integrals of the form
$\int Y^*_{l'm'} Y^{}_{1m''} Y^{}_{lm}$ \cite{arfk85}.

\section{Structure Optimization}\label{Smini}
We are now in a position to minimize the total
energy $\Et(\R)$ of a system with $M$ independent
atoms with respect to $3M$-dimensional position vector
$\R=(\R_1,\R_2,...,\R_M)$ using the directly
calculated force $\F= - d\Et/d\R$.
The simplest minimization scheme is to choose the
next geometry step always along the force direction
(steepest descent).
This method can be inefficient if the Born-Oppenheimer
surface happens to be a long and narrow valley.
In order to avoid oscillations within such a valley one
should take the previous minimization history into account.
This is, for example, accomplished by using one of the following
two procedures, the variable metric method or the damped Newton
dynamics scheme.

\subsection{Variable Metric Method}
If the total energy surface close to a geometry $\{\R^*\}$
is well-described by a quadratic approximation
\be\label{etot}
\Et(\R)= \Et(\R^*) - \F(\R^*) \cdot (\R-\R^*)
+ \frac{1}{2}(\R-\R^*) \cdot \A(\R^*) \cdot (\R-\R^*)\quad,
\ee
where $\A(\R^*)$ is the Hessian matrix, the variable metric
or quasi-Newton method provides a very efficient minimization.
The derivative of eq.\,(\ref{etot}) with respect to $\R$
leads to the following expression for the force $\F(\R)$
\be
\F(\R)= \F(\R^*) - \A(\R^*)\cdot(\R-\R^*)\quad .
\ee
Looking for the minimum of $\Et(\R)$ means searching for a zero
of this force. Hence, we have
\be\label{R-Ri}
\Delta \R^* \equiv \R-\R^*= \A^{-1}(\R^*)\cdot\F(\R^*)\quad.
\ee
The left side describes the finite step $\Delta \R^*$
which points into the minimum provided
the inverse Hessian $\A^{-1}(\R^*)$
and quadratic approximation of $\Et(\R)$ are exact.
Usually, this is not the case and
we face two serious problems: An exact inverse hessian is not
available and $\Delta \R^*$ from eq.\,(\ref{R-Ri}) may not direct
us in a downhill direction if higher order terms dominate the
description of the energy surface.

Fortunately, these difficulties can be removed by applying an
algorithm developed by Broyden, Fletcher, Goldfarb, and Shanno (BFGS)
\cite{broy73,brod77,denn83}.
Its realization within the variable metric method
as a FORTRAN program is described in Ref.\,\cite{pres92}.
The method iteratively builds up an approximation of $\A^{-1}(\R^*)$ by
making use of the forces obtained during previous steps
of the structure optimization.
This is done in such a way that the matrix remains positive
definite and symmetric.
This guarantees that $\Et(\R)$ decreases initially as we move into
the direction $\Delta \R^*$.
So, if the attempted step leads to an increase of the total energy,
i.e., $\Delta \R^*$ is too large, one just has to backtrack trying
smaller steps along the same direction in order to
obtain a lower energy.
The minimization process terminates
when {\em all} atomic forces for a geometry fall below
a certain limit.

\subsection{Damped Newton Dynamics and Molecular Dynamics}
The variable metric method works well if the energy surface
description is dominated by quadratic terms, e.g. close to a
total energy minimum.
However, if the quadratic approximation in eq.\,(\ref{etot})
is not well founded an algorithm based on damped Newton
dynamics is more robust and efficient \cite{stum94}.
In our approach we use for the time evolution of an arbitrary atomic
coordinate $R_m$ the finite difference equation
\be
R_m^{\tau+1}= R_m^\tau + \eta_m (R_m^\tau
- R_m^{\tau-1})+\delta_m F_m^\tau
\ee
where $R_m^\tau$ and $F_m^\tau$ are the coordinate and
the respective force at time step $\tau$.
Note that the minimization always includes implicitly the
history of displacements stored in the ``velocity''
coordinate $(R_m^\tau-R_m^{\tau-1})$.
Damping and speed of motion are controlled by the two
parameters $\eta_m$ and $\delta_m$.
An optimum choice of these two quantities would provide for
a fast movement towards the closest local minimum on the
Born-Oppenheimer surface and suppress oscillations around
this minimum.
A small damping factor $\eta_m$ improves the stability
of the atomic relaxation while a larger value allows
for energy barriers to be overcome and thus to escape
from local minima.
Again, the relaxation continues until all force components
are smaller in magnitude than a certain limit.
Obviously, if the damping is switched off, i. e. $\eta_m=1$
the approach equals an {\em ab-initio} molecular dynamics method.

\section{Examples}
In the following we present two test cases, the free H$_2$ molecule
and the hydrogen atom as an adsorbate on the (110) surface of bcc Mo.
The purpose of our study here is mainly to point out the agreement
between our forces and {\em numerical} derivatives of the total energy.
Also, we like to demonstrate the efficiency of our structure optimization.

The free H$_2$ molecule in the first example is
modeled in a cubic unit cell with a side length of 10\,bohr.
For the xc-potential we employ the generalized gradient
approximation (GGA) \cite{perd92}.
We choose a MT radius of 0.65\,bohr, and for the LAPW basis set
we use radial functions in the MT spheres up to $l_{\rm max}=8$
and a plane wave basis expansion in the interstitial region up
to $(K^{\rm wf})^2= 12$\,Ry.
The $(l,m)$ expansion for the potential goes up to $l_{\rm max}=4$.
Because of the small hydrogen MT radius a relatively high plane-wave
cutoff energy $E^{\rm cut,pot}=\left(G^{\rm pot}\right)^2=169$\,Ry
is necessary in
order to obtain a converged interstitial representation of the
potential. In Fig.\,\ref{dimer.H} we present directly calculated
\bF
\psfig{figure=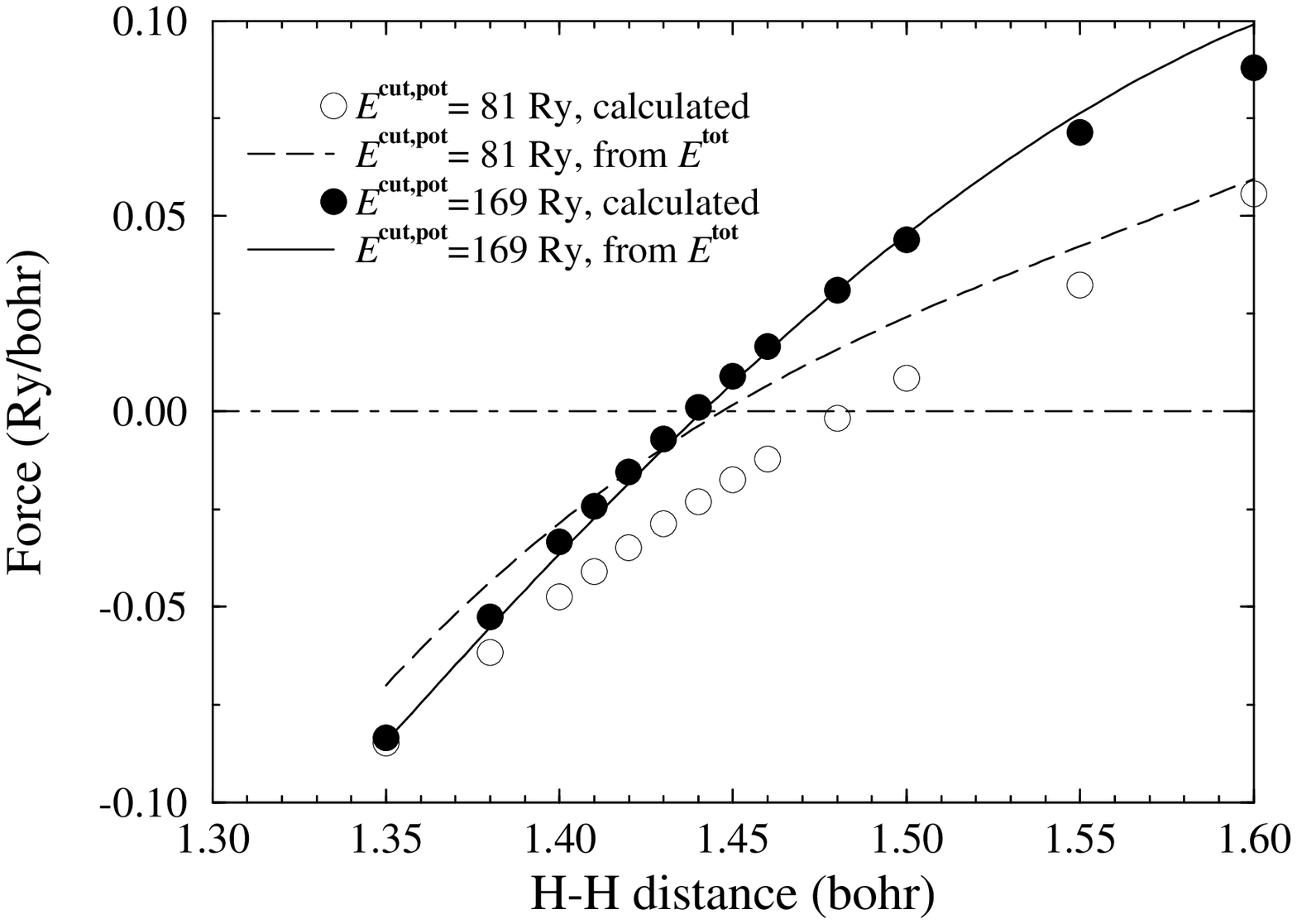,height=10cm}
\caption{\label{dimer.H} Atomic forces for a H-dimer as
a function of the H-H interatomic distance.
The data points represent the directly calculated forces
on the H-atom while the lines stem from polynomial fits to the total
energy.}
\eF
forces (solid dots) and compare them to forces obtained from a
polynomial fit to the respective total energies (solid lines). The
results demonstrate the excellent agreement between both data sets. If
the potential cutoff is too small ($E^{\rm cut,pot} = (G^{\rm pot})^2
= 81$\,Ry) the directly (empty dots) and indirectly (dashed lines)
evaluated forces differ considerably from each other.
To our knowledge no other element besides hydrogen exhibits such a
high sensitivity of the calculated atomic forces to the interstitial
representation of the potential.
As will be shown in the second example the $G^{\rm pot}$ cutoff
parameter is also less critical for hydrogen if a larger MT radius
can be chosen.

As a second case we present the example of a
structure optimization using the BFGS-minimization algorithm.
The goal is to find the relaxed geometry of the Mo\,(110) surface
covered with a full monolayer of hydrogen.
This problem is of particular interest because it was suspected
that the hydrogen adsorption induces a
so-called top-layer-shift reconstruction, i.e., a
shift of the Mo surface layer along the $[\overline{1}10]$
direction relative to the bulk \cite{altm87}.
The substrate surface is modeled by a five layer slab repeated
periodically and separated by 16.6\,bohr of vacuum.
Details of the geometry are shown in Fig.\,\ref{slab.HMo110}.
\bF
\psfig{figure=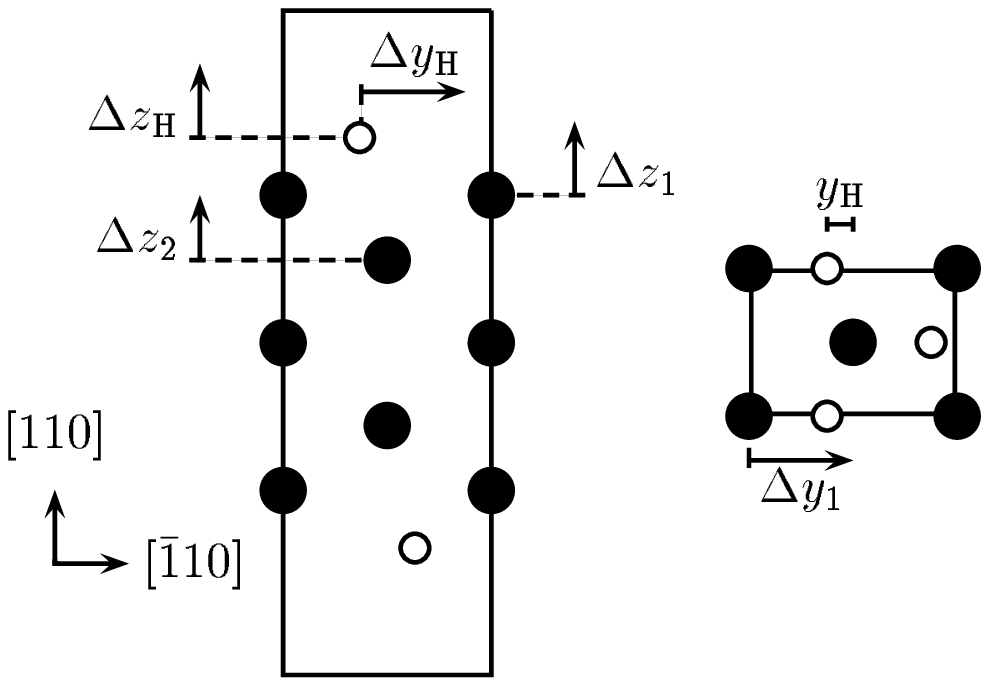,height=10cm}
\caption{\label{slab.HMo110}
Side (left) and top view (right) of the fully relaxed
5-layer H/Mo\,(110) system. The solid and empty circles
represent Mo- and H-atoms, respectively.}
\eF
The calculated in-plane lattice constant of 5.91\,bohr is used.
The $\vec{k}$-integrations are evaluated on a mesh of 64 equally
spaced points in the surface Brillouin zone.
The MT radii are chosen to be 2.40\,bohr and 0.90\,bohr for Mo
and H, respectively.
The kinetic-energy cutoff for the plane wave basis needed for
the interstitial region is set to 12\,Ry, and the $(l,m)$
representation (inside the MTs) is taken up to $l_{\rm max}= 8$
for both Mo and H.
Here, the hydrogen MT radius is relatively large compared to
that used for the study of the H-dimer.
Therefore, a plane-wave cutoff energy of 64\,Ry~for the
representation of the potential is sufficient.
The maximum angular momentum of the MT $(l,m)$ expansion of
the potential is set to $l_{\rm max}= 4$.
All states are treated non-relativistically.

In the beginning adsorbate, surface, and subsurface atoms are
distorted along the $[\overline{1}10]$ direction with respect
to the clean (110) surface configuration in order to break
the mirror symmetry of the clean surface.
During the relaxation process visualized in Fig.\,\ref{relax.HMo110}
\bF
\psfig{figure=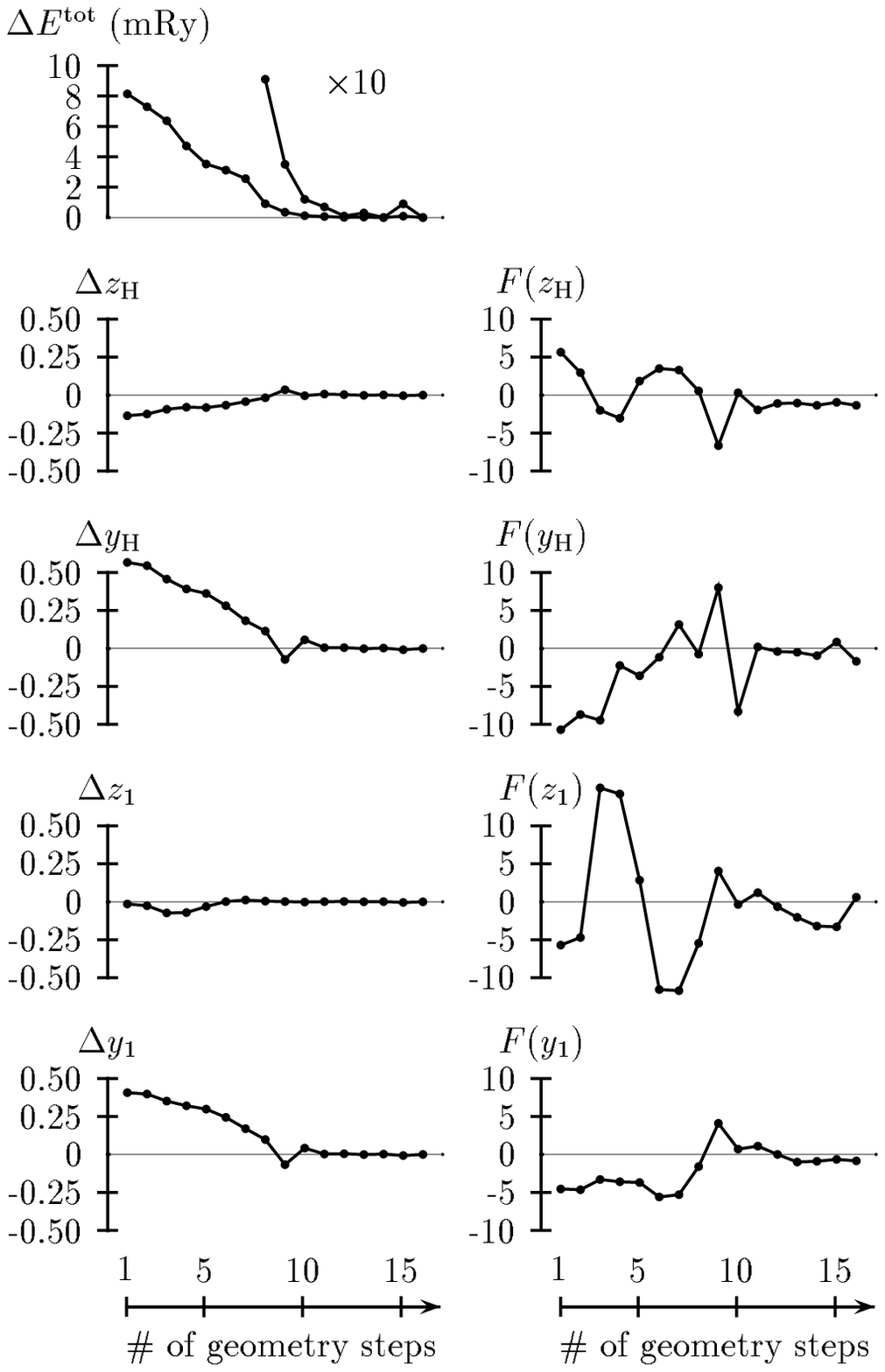,height=18cm}
\caption{\label{relax.HMo110}
H-Adsorption on Mo(110):
Minimization of the total energy for a slab with five Mo-layers.
Shown are the distances (in bohr) of the adsorbate and surface
atom positions from the optimized structure (left).
The respective forces (right) are given in mRy/bohr.
The structure parameters $y_{\rm H}$, $z_{\rm H}$, $y_1$, and $z_1$
are defined in Fig.\,\protect{\ref{slab.HMo110}}. }
\eF
all atoms are allowed to move freely perpendicular to the
surface and parallel along the $[1\bar{1}0]$-direction.
The system is considered to be in a stable or (metastable)
geometry when all force components are smaller than $3$\,mRy/bohr.
The error in the structure parameters of the relaxed system is
$\pm 0.02$\,bohr.

During the optimization process
the hydrogen relaxes into a quasi threefold position with a
$[\overline{1}10]$-offset of $y_{\rm H}= 1.19$\,bohr
from the long-bridge position and a height
of $d_{\rm H-Mo}=2.04$\,bohr above the surface layer.
For the surface layer we find a relaxation of $3\pm 0.3\,\%$ of the
bulk inter-layer spacing.
Furthermore, the surface atoms are slightly distorted by $0.09$\,bohr
along the $[\overline{1}10]$-direction opposite to the hydrogen
with respect to the subsurface.
Thus, there is no evidence for a pronounced top-layer-shift
reconstruction.
The subsurface layer relaxes to its bcc lattice position.

\section{Structure of the Program}
\subsection{Force Calculation}
The force implementation \verb|fhi95force| was developed for
\verb|WIEN|  \cite{blah90} as well as for
\verb|WIEN93| \cite{blah93a} which is a considerably
improved version of the original package.
Furthermore, our program is now part of the latest update
called \verb|WIEN95| \cite{blah95}.
Our intention was to keep the changes related to the
original code and the input-files as small as possible.
The program and running structure of the program were
kept unchanged.
In the following, we summarize the basic features of
the force computation within \verb|WIEN|:
\bi
\item[$\F_I^{\rm HF}$]: \verb|lapw0| \\
The electrostatic potential  is determined
on a logarithmic radial mesh.
For the calculation of $\F_I^{\rm HF}$
according to eq.\,(\ref{fhf}) we use the first radial mesh
point $r_{I,0}$.
Our experience is that $\F_I^{\rm HF}$ behaves
numerically stable close to the MT center, i.e., one
could also choose $r_{I,1}$ or $r_{I,5}$.
Also, the influence of the logarithmic radial mesh
chosen is uncritical.
The resulting Hellmann-Feynman force $\F_I^{\rm HF}$ is written
to unit 70 (\verb|case.fhf|).

\item[$\F_I^{\rm val}$]: \verb|lapw1| \& \verb|lapw2|\\
The determination of the correction $\F_I^{\rm val}$
is almost completely done in the subroutine \verb|l2for| (of \verb|lapw2|)
which resembles the original \verb|WIEN|-subroutine \verb|l2main|
plus the FORTRAN translation of the YSR eqs.\,(A12), (A17),
and (A20).
For the evaluation of $\int d\r^3 \rho(\r)\nabla V(\r)$
the subroutine \verb|fvdrho| is called.
The only additional input needed is related to the
non-spherical matrix elements
$\langle u_{l'}|V_\nu| u_l\rangle$,
$\langle u_{l'}|V_\nu| \dot{u}_l\rangle$,
$\langle \dot{u}_{l'}|V_\nu| u_l\rangle$, and
$\langle \dot{u}_{l'}|V_\nu| \dot{u}_l\rangle$
in eq.\,(A20) of YSR.
They are calculated in program \verb|lapw1| as a part of
the hamiltonian setup and transferred to \verb|lapw2| using
the input/output unit 71 (\verb|case.nsh(s)|).
The final result $\F_I^{\rm val}$ is written to unit 72
(\verb|case.fval| or \verb|case.fsc|).
The force calculation which is only executed if the switch \verb|FOR|
instead of \verb|TOT| is set in \verb|case.in2(s)| roughly triples
the running time of \verb|lapw2|.

\item[$\F_I^{\rm core}$]: \verb|core|\\
The core correction in eq.\,(\ref{fcor}) is calculated by the
routine \verb|fcore| using eq.\,(\ref{int}).
Before that the non-spherical potential has to be read from unit\,19
\verb|case.vns|. Note that $\rho_{\rm core}(\r)$ is spherical.
Therefore, only the effective potential parts
$V_{lm}^{\rm eff}(r)$ with $l=1$ have to
be taken into account.
The subroutine \verb|fcore| uses unit 73 (\verb|case.fcor|)
for the output of $\F_I^{\rm core}$.

\item[$\F_I$]: \verb|mixer|\\
The final step which is the summation of all (available) partial
forces according to eq.\,(\ref{fa}) is done by \verb|mixer|.
This executable writes the  accumulated information to unit 70
(\verb|case.ftot|) and unit 80 (\verb|case.finM|). The latter
can be used as input file for the minimization program \verb|mini|.
\ei
Note that all forces written to the output are in Ry/bohr.
Only the calculation of $\F_I^{\rm val}$
can be switched on and off the reason being that
the computer time due to $\F_I^{\rm HF}$,
$\F_I^{\rm core}$, and the output of the
non-spherical matrix elements in \verb|lapw1| is negligible.
The partial forces $\F_I^{\rm HF}$ and $\F_I^{\rm core}$  are
also excellent indicators to monitor the convergence of $\F_I^{\rm val}$
as well as the total atomic force $\F_I$.
Therefore, it is convenient to run a self-consistent calculation
until $\F_I^{\rm HF}$ or $\F_I^{\rm core}$ are converged and then evaluate
$\F_I^{\rm val}$ as a final step.
In this way the additional computing time necessary for the
atomic forces can be kept at a very low level.

\subsection{Minimization}
The program \verb|mini| is executed by invoking the UNIX command
`\verb|mini < mini.def|' in the case-directory.
As a first step, the minimization option and the control parameters
are read from unit\,5 (see Table\,\ref{TinM}).
\begin{table}{\footnotesize\tt
\begin{center}
\begin{tabular}{lp{2cm}l}
\hline
NEWT            &&minimization modus: BFGS/NEWT\\
0.7  2 2 0    &&eta, delta(1-3) of atom 1\\
0.7  1 1 1    &&eta, delta(1-3) of atom 2\\
... &&\\
\hline
\end{tabular}
\end{center}
}\caption{\label{TinM}
Input file unit 5 ({\tt case.inM})
}\end{table}
Then the previous minimization history from unit\,16
(\verb|case.tmpM|) is used to internally generate the
approximate inverse Hessian matrix (the velocity coordinate)
if the BFGS formalism (damped Newton dynamics scheme) is employed.
Now, the calculated total energy and forces are
read from unit\,15 (\verb|case.finM|) and used to determine the
next trial step.
If the new geometry causes an overlap of MT spheres a smaller
step along the same direction leading to touching MTs
is chosen.
Finally, the new trial geometry as well as the updated
minimization history are written to unit\,21 (\verb|case.struct1|)
and unit\,16 (\verb|case.tmpM|), respectively.

At this point, the FP-LAPW calculation using \verb|WIEN| for the new
geometry can take place.
The procedure continues until no further energy minimization
can be achieved or all atomic forces fall below a certain minimum.

\section{Installation}

The program package \verb|fhi95force| is written in FORTRAN77 and
should work on all machines where \verb|WIEN| can be installed.
Because changes of the original code have been kept at a minimum the
adaptation of \verb|fhi95force| to other versions of \verb|WIEN|
and to similar FP-{}LAPW codes should cause no problems.
The extension and the data files are contained in a single \verb|tar|
archive called \verb|fhi95force.tar|. The extraction on a UNIX
machine should generate a directory entitled \verb|fhi95force|
with the following subdirectories:
\bi
\item
\verb|SRC_force|\\
This directory contains the source files of the subroutines
which have to be
added to the existing code, e.g. \verb|lapw0_f.f| is the
extension for \verb|lapw0|.
(Sub)routines marked with an asterisks in Table\,\ref{Tsource}
\begin{table}{\footnotesize\tt
\begin{center}
\begin{tabular}{|l|ll|p{9cm}|}
\hline\hline
\rm source &&\rm routine &\rm relevance for the force calculation \\
\hline lapw0\_f.f
&*& lapw0
&\rm calculates and writes Hellmann-Feynman force to unit\,70\\
\hline lapw1\_f.f
&*& atpar &\rm writes non-spherical matrix elements to unit\,71\\
\hline lapw2\_f.f
&& charg2 &\rm modification of \verb|charge| \\
&& dfrad &\rm obtains radial derivative of a radial function\\
&*& lapw2 &\rm calls \verb|l2for| if switch \verb|FOR| is set\\
&& l2for &\rm evaluates and writes valence (semi-core) partial
forces to unit\,72\\
&& mag   &\rm obtains magnitude of a 3-dim vector\\
&& sevald &\rm computes first derivative of a spline\\
&& spline &\rm obtains coefficients for cubic interpolation spline\\
&& vdrho &\rm calculates $\int d^3\r V(\r) \nabla \rho (\r)$ \\
\hline core\_f.f
&& charge &\rm does Simpson integration inside a sphere\\
&& dfrad &\rm see \verb|lapw2_f.f|\\
&& fcore &\rm calculates core correction of force and writes result
to unit\,73\\
&*& hfsd &\rm calls \verb|fcore| \\
&& sevald &\rm see \verb|lapw2_f.f|\\
&& spline &\rm see \verb|lapw2_f.f|\\
\hline mixer\_f.f
&*& mixer &\rm calls \verb|totfor| \\
&& totfor &\rm calculates and writes total force to unit\,70\\
\hline mini.f
&&dfpmin &\rm minimizes a multi-dimensional function
using the BFGS variable metric method\\
&&finish&\rm writes final output\\
&&func	&\rm reads total energy and atomic forces from unit\,15\\
&&haupt &\rm handles input and output; calls \verb|dfpmin| or \verb|nwtmin|\\
&&inv	&\rm evaluates inverse of a matrix\\
&&latgen &\rm defines lattice (basis vectors)\\
&&lnsrch &\rm searches for a lower value of a multi-dimensional
function along the search direction \\
&&maxstp &\rm determines maximum possible atomic displacement without
overlap of MTs\\
&&nwtmin &\rm minimizes total energy using damped Newton dynamics\\
&&pairdis&\rm calculates pair distance of atoms\\
&&rotate &\rm rotates a vector\\
&&rotdef &\rm selects symmetry operations of equivalent atoms\\
\hline
\end{tabular}\end{center}
\vspace*{1cm}
}\caption{\label{Tsource}
Summary of the source files contained in the directory {\tt SRC\_force}.
The routines marked with an asterisk already exist in {\tt WIEN} and
have to be updated or substituted by the new versions.
}\end{table}
already exist in the original \verb|WIEN| code and
have to be removed before compiling the program.
If the local version of \verb|WIEN| differs from
the original one it should be very easy to update
the existing \verb|WIEN|-routines (\verb|lapw0|,
\verb|atpar|, \verb|lapw2|, \verb|hfsd|,
and \verb|mixer|) by hand.
The (few) changes necessary for the force calculations
are framed by FORTRAN comment lines (\verb|cfb| at the beginning
and \verb|cfe| at the end).
The compilation with \verb|make| can be done using
existing makefiles which, nevertheless, have to be updated with
the additional source files.
Before starting a calculation with the new executables it
will also be necessary to update the input/output channels
according to Table\,\ref{Tinout}.
\begin{table}
{\footnotesize
\begin{center}
\begin{tabular}{|l|rcl|p{7cm}|}
\hline\hline
\rm executable &\rm unit & I/O &\rm file-name &\rm comment \\
\hline \verb|lapw0|
& 70& O& \verb|case.fhf| &\rm Hellmann-Feynman force\\
\hline \verb|lapw1|
& 71& O& \verb|case.nsh(s)| &\rm non-spherical matrix elements\\
\hline \verb|lapw2|
&  2& I& \verb|case.in2(s)| &
\rm input file (switch for force-calculation)\\
& 19& I& \verb|case.vns| &\rm non-spherical potential\\
& 71& I& \verb|case.nsh(s)| &\rm non-spherical matrix elements\\
& 72& O& \verb|case.f[val/sc]| &\rm partial forces due to valence (semi-core)
electrons \\
\hline \verb|core|
& 19& I& \verb|case.vns| &\rm non-spherical potential\\
& 73& 0& \verb|case.fcor| &\rm partial force due to core electrons\\
\hline \verb|mixer|
& 70& O& \verb|case.ftot| &\rm sum of all available partial forces\\
& 71& I& \verb|case.fhf| &\rm Hellmann-Feynman force\\
& 72& I& \verb|case.fval| &\rm partial force due to valence electrons\\
& 73& I& \verb|case.fsc| &\rm partial force due to semi-core electrons\\
& 74& I& \verb|case.fcor| &\rm partial force due to core electrons\\
& 80& O& \verb|case.finM| &\rm input file for \verb|mini|\\
\hline \verb|mini|
& 5& I& \verb|case.inM| &\rm input-file\\
& 6& O& \verb|case.outputM| &\rm general output-file\\
&15& I& \verb|case.finM| &\rm total energy and atomic forces for current
geometry\\
&16& I/O& \verb|case.tmpM| &\rm history of minimization \\
&20& I& \verb|case.struct| &\rm current \verb|struct|-file \\
&21& O& \verb|case.struct1| &\rm \verb|struct|-file with new trial geometry\\
\hline
\end{tabular}\end{center}\vspace*{1cm}}
\caption{\label{Tinout}
Input and output-files relevant for the
force calculation and the minimization procedure.}
\end{table}
\item \verb|SRC_force93|\\
This is the \verb|WIEN93|-{}version of \verb|SRC_force|.
\item \verb|SRC_mini|\\
Here the source for the minimization program
can be found. Calling the UNIX command
\verb|make| within the directory creates the executable \verb|mini|.
\item
\verb|Mo|\\
This case directory enables the study of the H-{}point
phonon of bcc-{}Mo similar to the frozen-{}phonon
calculations presented in \cite{yu91}.
It can also be used for testing the minimization program
\verb|mini|.
\ei

\section{Test run}
We recommend to run the \verb|Mo| test case at the beginning with the
already existing local version of \verb|WIEN|, check the output for
warnings and error messages, and than repeat the procedure with the
updated \verb|WIEN|-{}version setting the \verb|FOR| switch
in \verb|Mo.in2| and \verb|Mo.in2s|.

Then, a step-by-step minimization running alternately
\verb|WIEN| and the new program \verb|mini| can take place.
If this one-dimensional relaxation leads successfully from
the distorted to the equilibrium bcc configuration, one may
go on to structure optimizations with higher
dimensionality.

\section{Acknowledgments}
This work has benefited from collaborations within, and has
been partially funded by, the Network on ``Ab-initio
(from electronic structure) calculation of complex processes
in materials'' (contract: ERBCHRXCT930369).
We thank P.~Blaha, P.~Dufek, and K.~Schwarz for fruitful discussions
and for providing us with the updated version \verb|WIEN93| of the
\verb|WIEN| code.
Furthermore, we like to acknowledge useful comments by
M.~F\"ahnle and contributions of G.~Vielsack to the
geometry optimization program.

{\renewcommand{\,}{$\!$\ }}
\pagestyle{empty}
\clearpage\newpage

\clearpage\newpage

\clearpage\newpage

\clearpage\newpage
\section*{Test run}
\subsection*{Output file \verb|Mo.force| (starting configuration)}
{\footnotesize
\begin{verbatim}
---------
  1.CYCLE
---------
 ja mu      F            |F|             Fx             Fy             Fz

  1  1    >>>   .0000000E+00   .0000000E+00   .0000000E+00   .0000000E+00
---------
  2.CYCLE
---------
 ja mu      F            |F|             Fx             Fy             Fz
  1  1    H-F   .3093266E+01   .0000000E+00   .0000000E+00  -.3093266E+01
  1  1    VAL   .3457794E-01   .8271806E-24   .3308722E-23   .3457794E-01
  1  1    VAL   .1951423E+00  -.3642919E-16  -.5637851E-17   .1951423E+00
  1  1    COR   .2423779E+01   .0000000E+00   .0000000E+00   .2423779E+01

  1  1    >>>   .4397668E+00  -.3642919E-16  -.5637848E-17  -.4397668E+00
...
---------
 11.CYCLE
---------
 ja mu      F            |F|             Fx             Fy             Fz
  1  1    H-F   .1621001E+01   .0000000E+00   .0000000E+00  -.1621001E+01
  1  1    VAL   .1334964E-01   .1654361E-23   .6203855E-24   .1334964E-01
  1  1    VAL   .6451487E-01  -.3089976E-17  -.4201283E-16   .6451487E-01
  1  1    COR   .1409447E+01   .0000000E+00   .0000000E+00   .1409447E+01

  1  1    >>>   .1336895E+00  -.3089974E-17  -.4201283E-16  -.1336895E+00
---------
 12.CYCLE
---------
 ja mu      F            |F|             Fx             Fy             Fz
  1  1    H-F   .1621459E+01   .0000000E+00   .0000000E+00  -.1621459E+01
  1  1    VAL   .1335570E-01  -.4135903E-24  -.1240771E-23   .1335570E-01
  1  1    VAL   .6455603E-01   .3474868E-16  -.9199455E-16   .6455603E-01
  1  1    COR   .1409757E+01   .0000000E+00   .0000000E+00   .1409757E+01

  1  1    >>>   .1337903E+00   .3474868E-16  -.9199455E-16  -.1337903E+00
\end{verbatim}}

\clearpage\newpage
\subsection*{Output file \verb|Mo.tmpM| (after three geometry steps)}
{\footnotesize
\begin{verbatim}
  3
    3
           -93.75512         1
   .147750000000D+01   .000000000000D+00         1.4775     .0000     .2500
   .147750000000D+01   .000000000000D+00         1.4775     .0000     .2500
   .159570000000D+01   .668951500000D-01         1.5957     .0669     .2700
           -93.76944         2
   .147750000000D+01   .000000000000D+00         1.4775     .0000     .2500
   .147750000000D+01   .000000000000D+00         1.4775     .0000     .2500
   .152880471000D+01   .284468350000D-01         1.5288     .0284     .2587
           -93.77203         3
   .147750000000D+01   .000000000000D+00         1.4775     .0000     .2500
   .147750000000D+01   .000000000000D+00         1.4775     .0000     .2500
   .145353140400D+01  -.133528500000D-01         1.4535    -.0134     .2459
          .0000000         4
   .147750000000D+01   .000000000000D+00         1.4775     .0000     .2500
   .147750000000D+01   .000000000000D+00         1.4775     .0000     .2500
   .144430222800D+01   .000000000000D+00         1.4443     .0000     .2444
\end{verbatim}}
\end{document}